\documentclass{article}
\usepackage{spconf,graphicx, cite}
\usepackage{float}
\usepackage{ragged2e}
\usepackage{enumitem}
\usepackage[caption=false]{subfig}
\usepackage{pifont}
\usepackage{amsmath,amssymb} 
\usepackage[linesnumbered,ruled,vlined]{algorithm2e}
\usepackage[table, svgnames, dvipsnames]{xcolor}
\usepackage{bm}      
\usepackage{comment}

\title{Recent Advances in Scalable Energy-Efficient and Trustworthy Spiking Neural Networks: from Algorithms to Technology}
\name{\begin{tabular}{c}Souvik Kundu$^{1}$, Rui-Jie Zhu$^{2}$, Akhilesh Jaiswal$^{3}$, Peter A. Beerel$^{4}$\end{tabular}}
\address{$^{1}$Intel Labs, San Diego;  $^{2}$University of California, Santa Cruz\\$^{3}$University of Wisconsin-Madison; $^{4}$University of Southern California}
\begin{document}
%
\maketitle
\begin{abstract}
Neuromorphic computing and, in particular, spiking neural networks (SNNs) have become an attractive alternative to deep neural networks for a broad range of signal processing applications, processing static and/or temporal inputs from different sensory modalities, including audio and vision sensors. 
In this paper, we start with a description of recent advances in algorithmic and optimization 
innovations to efficiently train and scale low-latency, and energy-efficient 
spiking neural networks (SNNs) for complex machine learning applications. 
We then discuss the recent efforts in algorithm-architecture co-design that explores the 
inherent trade-offs between achieving high energy-efficiency and low latency while still
providing high accuracy and trustworthiness.
We then describe the underlying hardware that has been developed to leverage such algorithmic innovations in an efficient way. In particular, we describe a hybrid method to integrate significant portions of the model's computation within both memory components as well as the sensor itself.  Finally, we discuss the potential path forward for research in building deployable SNN systems identifying key challenges in the algorithm-hardware-application co-design space with an emphasis on trustworthiness.
\end{abstract}

\section{Introduction}
Despite the massive commercial success of deep neural networks (DNNs), the exponential growth of machine learning models and associated training and inference computational as well as carbon footprint has become alarming. Recent studies found that training a state-of-the-art (SOTA) artificial neural network (ANN) model can cost between \$3M and \$30M with projected costs reaching \$500M by 2030 and can create 626,000 pounds of planet-warming carbon dioxide, equal to the lifetime emissions of five cars \cite{strubell2019energy}. Moreover, inference platforms, particularly at the edge due to resource and energy constraints, have significant challenges keeping up with these increased model sizes \cite{chen2022mobile}. This has triggered significant research on various energy-efficient alternatives to traditional ANNs including Spiking Neural Networks (SNNs).
\begin{figure}[t!] 
    \centering
    \includegraphics[width=0.42\textwidth]{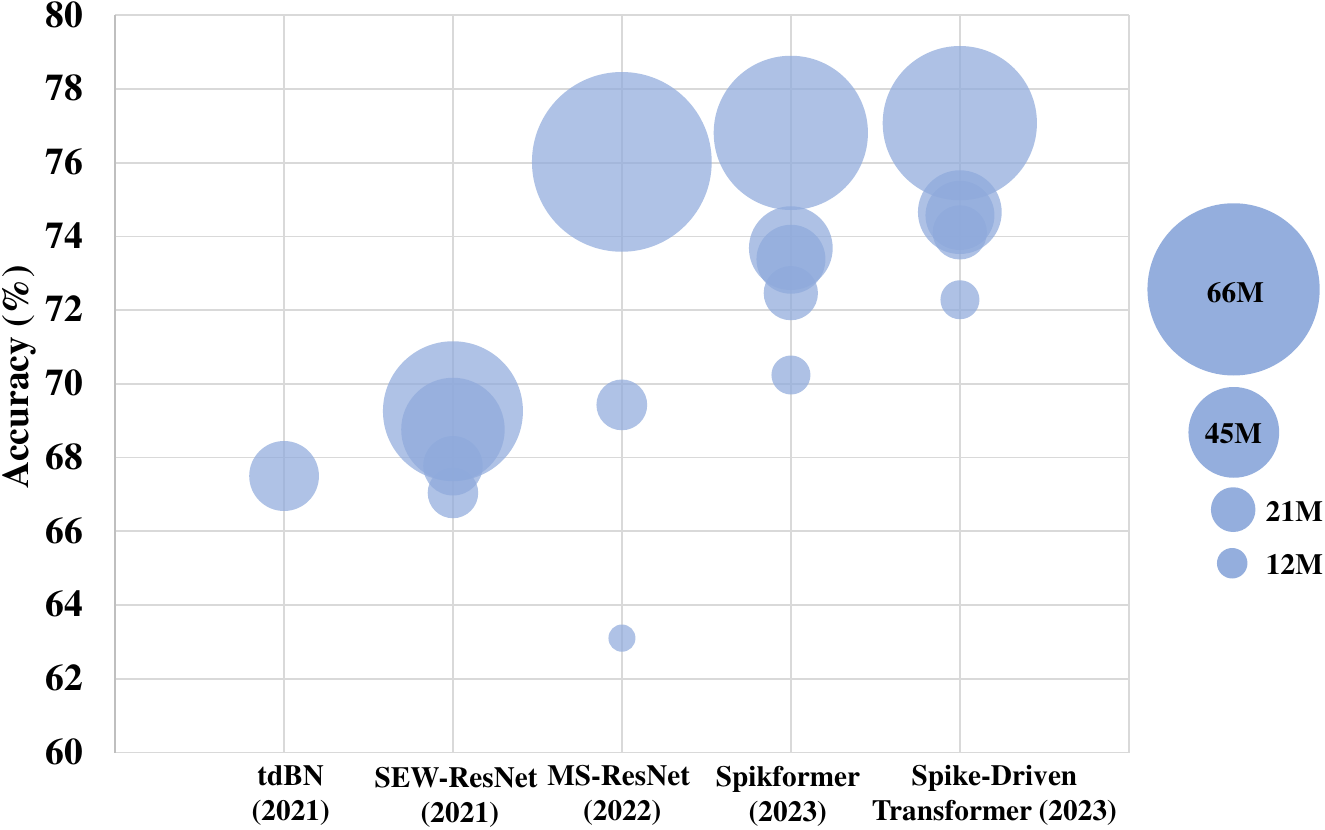}
    \vspace{-3mm}
\caption{Growth trend of deep SNN vision model sizes and test accuracy improvement on ImageNet in recent years, including tdBN~\cite{zheng2021going}, SEW-ResNet~\cite{fang2021deep}, MS-ResNet~\cite{hu2021advancing}, Spikformer~\cite{zhou2023spikformer} and Spike-Driven Transformer~\cite{yao2023spike}.}
    \label{fig:deep_snn_size_trend}
    \vspace{-4mm}
\end{figure}

While traditionally difficult to train, recently proposed supervised training algorithms have yielded SNNs that demonstrated high energy efficiency and accuracy for a wide range of models and applications, particularly for event-based sequential learning applications (e.g., \cite{Ponghiran2021SpikingNN}). The potential for SNN's energy efficiency stems from their spiking nature (see e.g. \cite{ottati2023spike}). In contrast to operating on multi-bit activation as in ANNs, SNNs typically operate on binary representation of the event-based spikes. This replaces the compute heavy multiply-accumulate (MAC) needed in ANNs with far more efficient conditional accumulation. Moreover, the training algorithms have helped limit the number of spikes by reducing the number of time steps needed to encode inputs \cite{datta2021training}, promoting sparsity within the network \cite{rathi2020iclr}, and increasing sparsity of spike generation \cite{kundu2021spike}. Coupled with advances in in-memory and in-sensor computing using both CMOS \cite{datta2023sensor} and next generation technologies \cite{Yang2022}, hardware accelerators that can take advantage of these features, the potential for energy-efficient SNN-based solutions is significant. 

These solutions are actively being expanded to consider other important factors including privacy as well as robustness to both adversarial and natural perturbations, particularly for safety-critical edge-based signal-processing applications \cite{Kundu_2021_ICCV}. The potential SNN applications include both static and temporal tasks processing data from a wide range of sensor modalities, e.g., audio, video, and bio-signals, for a variety of complex systems ranging from satellites to robots. Having said this, it is important to note that the huge success of ANNs in static tasks has created a steep learning curve for SNNs to catch. Fortunately, SNNs are inherently likely to be more suitable for event driven as well as inputs with temporal information. Also, the immense potential for low-power hardware to accelerate large-scale SNNs has motivated efficient and effective on-chip learning via SNNs \cite{ottati2023spike}.

\noindent
\textbf{Our contributions.} To better understanding the potential of SNNs, we survey recent advances in SNNs from algorithms to hardware acceleration. In particular, we discuss the latest advances in training approaches, spiking model development scaled up to million-scale, and in-sensor based SNN computing hardware. Fig. \ref{fig:deep_snn_size_trend} demonstrates the evolution of deep SNNs over the past few years showing an encouraging trend of improved classification accuracy. We further discuss the advancement and limitations on the trustworthiness of the large scale SNNs for safety-critical deployment. We review the latest trend in hardware for SNNs including analog and digital approaches spanning CMOS and emerging technologies and the need for hardware-algorithm co-design. Finally, we present discussion on various pressing problems that the research community may handle to make SNNs the de-facto alternative to ANNs for various real-life applications. 

\section{Scalable SNNs: Training and Models} 

\subsection{Training Towards Deep SNNs}
%
\begin{figure*}[!t]
    \centering
    \vspace{-4mm}
        \subfloat[]{\includegraphics[width=12.4cm]{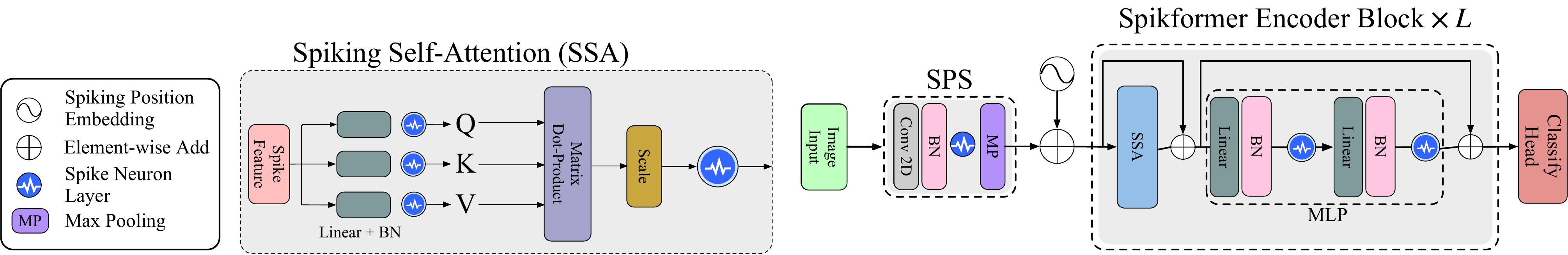}}\hspace*{0.5cm}
        \subfloat[]{\includegraphics[width=4.7cm]{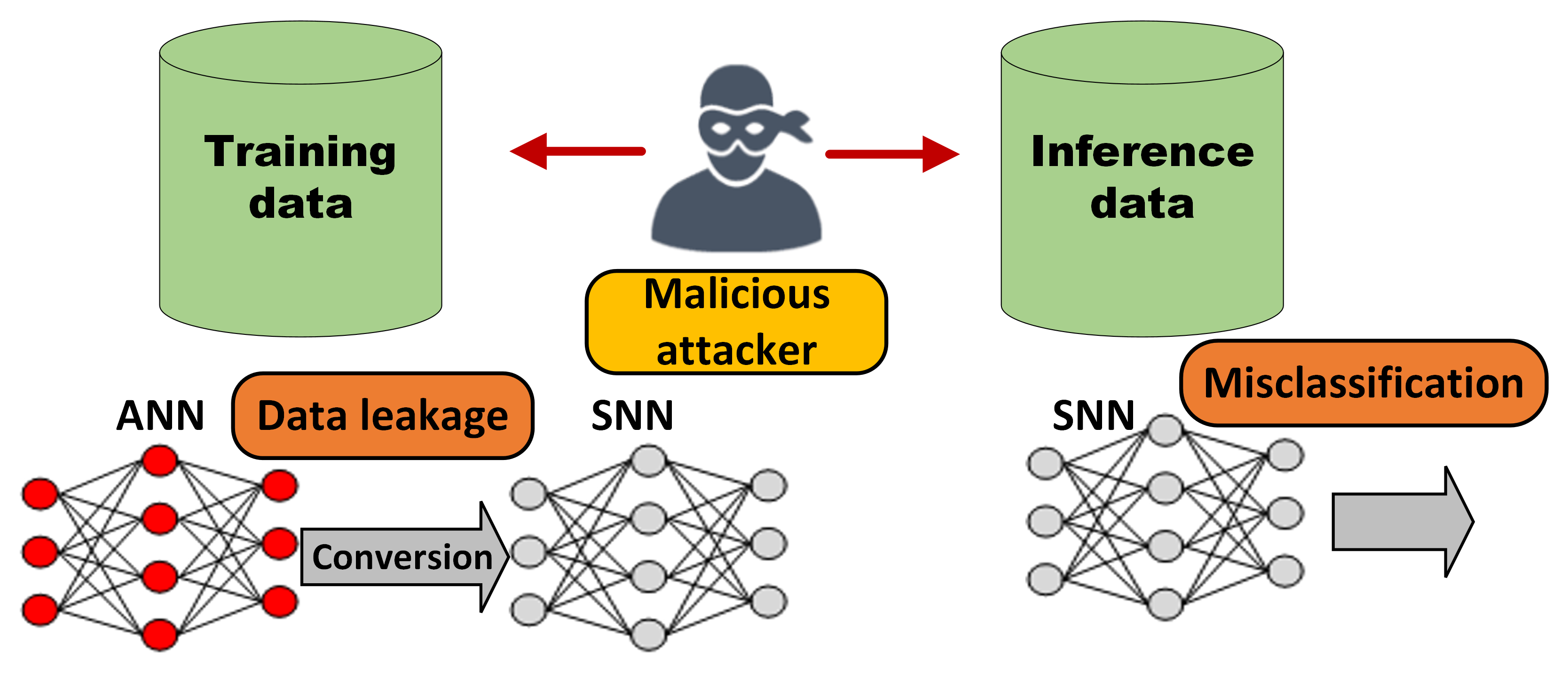}}
    \caption{(a) The spiking self attention based transformer architecture, (b) example of attack scenarios for SNNs.}
    \vspace{-4mm}
    \label{fig:spiking_transformer}
\end{figure*}

\noindent
\textbf{Input Encoding.} Over the years, researchers have proposed solutions to train spike-based training of SNNs for deeper models without loss of information, and yield close to the ANN accuracy. 
To handle multi-bit inputs, such as typical in real-life sensor-based applications, inputs are often spike encoded in the temporal domain using rate coding \cite{diehl2016conversion}, temporal coding \cite{comsa_2020}, or rank-order coding \cite{Kheradpisheh_2020}. Alternatively, some researchers advocate directly feeding the analog pixel values in the first 
convolutional layer, and thereby, emitting spikes only in the subsequent layers \cite{rathi2020dietsnn}. 
This can reduce the number of time steps needed to achieve the SOTA accuracy, 
however, comes at the cost that the first layer now requires MACs \cite{rathi2020dietsnn}, making  
hardware acceleration complex.

\vspace*{0.1cm}
\noindent
\textbf{Initialization and training.} Based on the initialization methods to yield faster and improved convergence, SNN training can be broadly classified into three categories, \textit{ANN-SNN conversion} \cite{sengupta2019going}, \textit{SNN training from scratch} \cite{lee2019enabling}, and \textit{hybrid training} \cite{rathi2020iclr}. The notion of back propagation through discrete time-steps has been enabled by various approximation of gradients for SNN training including STE \cite{sengupta2019going}. 
In particular, \cite{sengupta2019going} has demonstrated the efficacy of batch-normalization (BN)-less ResNet ANNs that are then converted in SNN domains. 
SNN training from scratch on the other hand yields lower time steps but are generally inferior in accuracy \cite{sengupta2019going}. 
The hybrid approach starts with a ANN-SNN converted model and then fine-tunes the model using back propagation for a few epochs to reduce the number of time steps needed. For training a model in the spiking domain, the neuron membrane dynamics can be modeled as
\vspace{-1.5mm}
\begin{align}
    u_i^{t+1} &= u_i^t + \sum_j m_{ij}*w_{ij}O_j^t - v_{th}O_i^t\\
   O_i^{t}   &=
    \begin{cases}
    1, & \text{if } z_i^{t}>0, \\
    0, & \text{otherwise}
    \end{cases} 
    \label{eq:sparse_snn_train}
\end{align}
\noindent
Here $z_i^{t}$ = ($\frac{u_i^t}{v_{th}} -1$) denotes the normalized membrane potential and $m_{ij} \in \{0,1\}$ denotes the fixed sparsity-mask between a neuron $i$ and its pre-synaptic neuron $j$, where a 0 and 1 indicate absence and presence of synaptic weights, respectively. For traditional non-sparse training, $m_{ij} = 1$.
During backpropagation, the weight gradient via activation spikes is
\begin{align}
    \delta w_{ij} = m_{ij} * \sum_t \frac{\partial\mathcal{L}}{\partial O_i^t}\frac{\partial O_i^{t}}{\partial z_i^{t}}\frac{\partial z_i^{t}}{\partial u_i^{t}}\frac{\partial u_i^{t}}{\partial w_{ij}^t} 
    \label{eq:sparse_grad_update}
\end{align}

\noindent
Here $O_i^t$ is the thresholding function. The term $\frac{\partial O_i^{t}}{\partial z_i^{t}}$ requires a pseudo-derivative as mentioned in \cite{bellec_2018long} and can be define as
\begin{align}
    \frac{\partial O_i^{t}}{\partial z_i^{t}} = \gamma * max\{0, 1-|z_i^{t}|\} 
    \label{eq:linear_surrogate}
\end{align}
\noindent

It is noteworthy, that the conversion method generally has low training cost due to the training being done in ANN domain without the need of back propagation through time (BPTT). This has motivated several recent works to focus on reducing the conversion loss in ANN-SNN conversion method by smartly approximating the activation value of ReLU neurons with the firing rate of spiking neurons \cite{deng2021optimal, li2022converting, dsnn_conversion_abhronilfin, rathi2020iclr, diehl2016conversion}.  
In particular, several works proposed shifting the ReLU activation functions in CNNs for 
identical and independent distributions (IID)~\cite{deng2021optimal,Bu_Ding_Yu_Huang_2022} and
skewed distributions \cite{datta2021training}, the latter proposing to learn the optimal shift 
during training.  
Moreover, researchers have proposed weight sparsification methods via spiking 
attention guided compression \cite{kundu2021spike, kundu2021towards} and gradient rewiring \cite{chen2021pruning} that can reduce parameter count by up to $\mathord{\sim}33\times$.

The resulting SNNs have been able to perform similar to SOTA convolution based ANNs in traditional 
image recognition tasks \cite{rathi2020dietsnn, yao2023attention} and dynamic vision tasks \cite{dynvis_1,dynvis_2}. 

\subsection{Beyond Spiking Residual Networks} 

Following the rise of SOTA convolution based ANN architectures like ResNet~\cite{he2016resnet}, the deep learning community has pivoted to adapt self-attention (SA) modules ~\cite{vaswani2017attention, kundu2021attentionlite} as the next de-facto module for ANNs. Similarly, building upon the accomplishments of traditional ANN Transformers, very recently, the spiking Transformer has emerged as a promising solution poised to take deep SNN architectures to the next level. The spiking transformer represents a promising direction for advancing neural networks through innovations like incorporating the spiking self-attention (SSA) module. We now discuss the advancement of the spiking transformer architecture for both natural language processing (NLP) and computer vision tasks.

\vspace*{0.1cm}
\noindent
\textbf{Spiking transformer for Vision.}
In the realm of vision tasks, the Spiking Transformer has demonstrated its prowess with models of up to 66 million parameters, outshining existing SNN models and even rivaling the performance of ANN vision transformer models of comparable size on benchmarks like ImageNet-1K~\cite{deng2009imagenet}. 
The Spikformer~\cite{zhou2023spikformer} pioneered the application of vision Transformers to spiking neural networks by incorporating three key innovations (Fig. \ref{fig:spiking_transformer}(a)). First, the fully-connected (FC) layer in the vanilla Transformer are replaced with FC-spiking neuron architecture. Second, Spikformer exploits the operation introduced in cosFormer~\cite{qin2021cosformer} which multiplies the transpose of key matrix $K^T$ with the value matrix $V \in n\times d$ to obtain a $d \times d$ attention map. Since $d << n$ in most models, this reduces the computational complexity from $O(n^2d)$ to $O(nd^2)$. Third, the SSA does not utilize the compute heavy softmax operation to normalize due to the inherent non-negative nature of the post query-key operation. Note, the attention map in spikformer uses integer rather than spiking representations. 

Subsequent efforts to optimize vision spiking Transformers have focused on the residual connections and attention maps. The Spikingformer~\cite{zhou2023spikingformer} replaces the spike residuals with membrane potential shortcuts, improving top-1 classification accuracy on ImageNet-1K by 1.04\%. The Spike-Driven Transformer~\cite{yao2023spike} uses binary spike-driven attention maps instead of integer maps and Hadamard products rather than matrix multiplications, achieving linear complexity while yielding 77.1\% top-1 accuracy on ImageNet-1K. Recently, \cite{zhang2022spiking} has presented a spiking transformer specifically dedicated for event-based object tracking, potentially targeting the optimal application space for SNNs while utilizing the benefits of attention mechanism.

\noindent
\textbf{Spiking transformer for NLP.} 
The application of spiking Transformers to NLP tasks remains relatively unexplored. To the best of our knowledge, SpikeGPT \cite{zhu2023spikegpt} is the only large scale model that has incorporated the idea of spiking self-attention in language modeling. To mitigate the quadratic computation complexity of attention mechanism, SpikeGPT builds upon the RWKV transformer~\cite{peng2023rwkv}, a linear complexity attention variant. 
Compared to vision spiking transformers, one of the fascinating aspects of the SpikeGPT architecture lies in its incorporation of spiking neurons, necessitating an additional dimension for the feed-forward process. In vision tasks, this is reminiscent of the successful approach taken by convolutional SNNs and vision spiking transformers, which integrate an extra temporal dimension to accommodate spiking neurons during feed-forward computations. Meanwhile, for NLP tasks, SpikeGPT aligns the spiking neuron feed-forward dimension with the sequence dimension, harnessing the intrinsic sequential temporal nature of language data to achieve optimal results. This intelligent architectural adaptation illustrates the elegance of the SpikeGPT and its ability to adapt to the distinct requirements of different language based inputs.

\section{Privacy and Robustness of Deep SNNs}
Along with the growing demand of energy-efficiency, the demand for trustworthiness both in terms of maintaining \textit{privacy} \cite{kundu2021analyzing, kundu2023learning} and \textit{robustness} \cite{Kundu_2021_ICCV} of the model has grown over the years (see Fig. \ref{fig:spiking_transformer}(b)). Deployment of SNNs in a trustworthy fashion while also preserving the benefits of energy efficiency has triggered research targeting efficient yet trustworthy algorithm design for neuromorphic computing. Here, privacy is associated to both model intellectual property (IP) as well as the data on which the model is trained. Robustness, on the other hand, can be attributed to the model's performance on out-of-distribution (OOD) dataset \cite{kundu2023sparse} that often belongs to the same class as the original training class, but generated via various malicious perturbations \cite{Kundu_2021_ICCV}.

\noindent
\textbf{Private SNNs.} Recently, few works \cite{kim2022privatesnn, kim2022privatesnn} have discussed  the potential of data and class leakage issue in SNN training via ANN-SNN conversion. In particular, to mitigate the leakage of the training data on which the ANN was trained, \cite{kim2022privatesnn} presented idea of performing conversion based training on synthetic data that can mimic the distribution of the original training set. Further to protect the model from class leakage researchers have proposed 
to encrypt SNN weights by training SNNs with a temporal spike-based learning rule. Additionally, research exploring SNN training in a federated environment has opened the door to perform training on event driven input at the user's local device without sharing their data~\cite{venkatesha2021federated}.

\noindent
\textbf{Robust SNNs.} Model's vulnerability on OOD data is a well known problem that persists for both ANNs and SNNs. \cite{sharmin2020inherent} argued that SNNs may be inherently robust due to their processing on spike based input helping them reduce the strength of the input pixel perturbations. Later \cite{Kundu_2021_ICCV} demonstrated that majority of the inherent robustness may be attributed to 
high-latency rate-coded input processing that are often absent in low-latency direct input coded SNNs. Additionally, \cite{Kundu_2021_ICCV} proposed a training method to improve the SNN inherent robustness by up to 10.1\% while evaluated on strong gradient based attacks, that too at ultra low latency applications with inference time steps as low as six.

\section{Devising the Technology for SNNs} 

Hardware computing platforms for accelerating SNNs broadly are driven by attempts to map characteristic processing aspects\cite{roy2019towards} - sparsity, distributed memory (synapses) and computing (neurons),  event-driven nature and temporal processing - of spike-based algorithms onto underlying hardware technology - CMOS or non-volatile memory (NVM) based digital or analog circuits. Additionally, researchers have been trying to adopt neuromorphic principles in to sensor design, specifically vision sensors \cite{lenero20113}, and more recently leveraging the emerging 3D heterogeneous integration technology \cite{iris} for embedding neuromorphic computing within sensors. 

Digital neuromorphic hardware like Intel's Loihi \cite{davies2018loihi} uses discrete time computational model to implement digital leaky-integrate-fire neurons using standard CMOS technology. Key neuromorphic aspects that have dictated Loihi's design are use of asynchronous circuits mapping the event-driven computing nature of neuromorphic algorithms, spike-event based messaging for communication through network-on-chip, configurable and distributed on-chip synaptic memory with routing look up tables and support for implementation of local learning rules like spike timing dependent plasticity (STDP). The major advantage of digital approaches like Loihi is the use of mature and robust CMOS technology leading to massive scalability of the hardware platform to multiple chips and modules and the availability of associated software and programming framework.

In contrast to digital solutions, analog neuromorphic hardware uses analog design techniques to implement continuous time computing models typically leveraging exponential characteristics of subthreshold CMOS circuits. Recently, the emergence of non-volatile memory technologies like resistive-RAM (RRAM) and phase change memory (PCM) have ushered in novel efforts to implement in-memory analog computing operations in an attempt to mitigate the overhead associated with the memory-wall bottleneck \cite{chakraborty2020pathways}. Interestingly, in-memory computing efforts have also been extended to CMOS based memory solution like SRAM \cite{jaiswal20198t} and present a promising pathway for use of mature CMOS technology for implementing neuromorphic computing with reduced overhead associated with the memory wall.

An emerging field in neuromorphic computing is the development of neuromorphic sensors. Specifically, advanced dynamic vision sensors have seen significant progress in embedding aspects of retinal processing into sensor \cite{lenero20113}, with recent works proposing to leverage 3D integration technology to map computations from photoreceptors to ganglion cells (retina output) into neuromorphic camera systems. 3D integration technology has also shown promise in integrating significant portions of SNN computations inside neuromorphic pixel array using 1:1 hybrid Cu-Cu bonding schemes \cite{iris}.

\begin{figure}[t!] 
    \centering
    \includegraphics[width=0.45\textwidth]{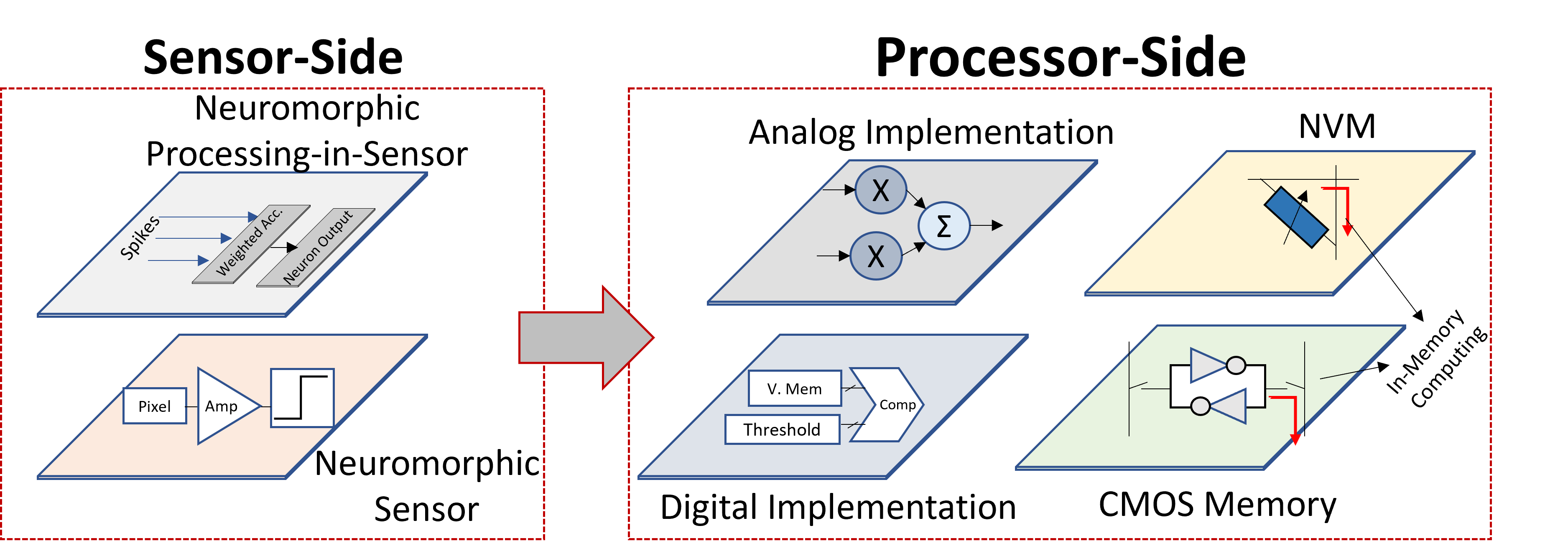}
    \vspace{-2mm}
    \caption{Overview of neuromorphic hardware approaches.}
    \label{fig:snn_hardware}
    \vspace{-4mm}
\end{figure}
\section{The Path Ahead for Deployable SNNs}

Based on the promise and potential limitations, we broadly identify four research thrusts for SNNs to emerge as a large scale deployable solution. 

\vspace*{0.1cm}
\noindent
\textbf{Scalability.} 
Despite the promising progress made by the spiking transformer, there remain two crucial challenges that need to be addressed: rich pre-training and ``billion-scale" scalability. 
Pre-trained models are one of the most vital features of standard ANN transformers~\cite{zou2023segment} as even relatively small-scale vision models leveraging pre-training can achieve strong generalization capabilities. However, pre-trained models have not yet been effectively explored for spiking  transformers.
Scalability is a feature of ANN transformer foundation models and the principles of scaling is a topic of extensive research  ~\cite{hoffmann2022training}. Although models like SpikeGPT have demonstrated scaling with up to 216 million parameters with pre-training, spiking transformers have not yet reached the largest ``billion-scale" ANN transformer models ~\cite{brown2020language,zeng2022glm}. 
This opens up an intriguing research branch of efficient self-supervised 
pre-training of large SNNs that can act as foundation models for various edge applications.

\vspace*{0.1cm}
\noindent
\textbf{Trustworthiness.} 
The second thrust is trustworthiness, demonstrating that SNNs cannot only be attractive for 
their energy efficiency, but can be robust to both naturally and adversarial perturbations
for a wide range of applications while maintaining privacy of the model and training data. This is important, as traditional ANN foundation models have shown significant vulnerability both in terms of privacy leaks and reduced robustness \cite{kundu2021dnr}. Only recently few works have \cite{kim2022privatesnn} shown promising results in understanding novel privacy threats in SNN. Developing of trustworthy SNN systems has thus remained an open yet interesting problem.
%

\vspace*{0.1cm}
\noindent
\textbf{Hardware.} 
Several intricate hardware-centric aspects are to be addressed to deliver a full-stack deployable neuromorphic system as a de-facto platform for specific niche applications. These hardware considerations are intricately tied to algorithmic implementations and include implementing aspects of input encoding on the hardware, interfaces to traditional computing platforms (CPUs and GPUs), circuit calibration specifically for analog platforms and including online training for adaptable and agile neuromorphic systems. Further, neuromorphic systems inevitably need to consider privacy and trustworthiness of the end application as an indispensable attribute of deployable systems. Thus, hardware-algorithm-application co-design is crucial for bringing neuromorphic systems into widespread practical use.

\vspace*{0.1cm}
\noindent
\textbf{Applications.} We have seen the great promise from SNNs for both static and event driven temporal input processing. However, it is imperative that traditionally the presence of time steps in the training method, makes the temporal input based applications of SNNs have an inherent advantage over the traditional ANNs. The success of recent research \cite{zhang2022spiking} on ultra-low power event based object tracking provides an encouraging future for SNNs, particularly for certain key applications. Based on this trend, we believe various temporal NLP and event driven low-power applications, are some of the promising directions for deep SNNs to be explored.

Challenges in all four areas are significant but advances are rapidly occurring and the specter of artificial intelligence  accelerating climate change has made this research, whether ultimately successful or not, a global imperative.


\bibliographystyle{IEEEbib}
\footnotesize
\bibliography{refs}

\end{document}